\documentclass[11pt]{article}

\usepackage[a4paper,margin=1in]{geometry}
\usepackage{graphicx}
\usepackage{amsmath, amssymb}
\usepackage{physics}
\usepackage{bm}
\usepackage{hyperref}
\usepackage{caption}
\usepackage{subcaption}
\usepackage{authblk}
\usepackage{xcolor}
\usepackage{float}

\hypersetup{
  colorlinks=true,
  linkcolor=black,
  citecolor=blue,
  urlcolor=blue
}

\title{Information Efficiency of Scientific Automation}

\author{Mihir Rao}
\affil{Princeton University}

\date{}

\begin{document}
\maketitle

\begin{abstract}
Scientific discovery can be framed as a thermodynamic process in which an agent invests physical work to acquire information about an environment under a finite work budget. Using established results about the thermodynamics of computing, we derive finite-budget bounds on information gain over rounds of sequential Bayesian learning. We also propose a metric of information-work efficiency, and compare unpartitioned and federated learning strategies under matched work budgets. The presented results offer guidance in the form of bounds and an information efficiency metric for efforts in scientific automation at large.
\end{abstract}

\section{Introduction}
Scientific research involves agents (people, labs, algorithms, or platforms) interacting with an environment to reduce epistemic uncertainty. When mediated by physical experiments, this process is constrained by the thermodynamic laws. Information thermodynamics provides rigorous lower bounds on the energetic cost of information acquisition \cite{Landauer1961,Bennett1982,Sagawa2010,Jarzynski2011,Toyabe2010,Still2012,Horowitz2014}. These bounds yield strategic implications for how to strategize automated science platforms. In this work, we use a simple model of \emph{measure-update-erase} cycles for information acquisition in automated science. We then offer a metric of information-work efficiency, which we believe to be positively correlated with a ``good'' automated science platform in practice, and comment on its regimes. Lastly, we derive the implications of partitioning a scientific domain, and the bounds of information-work efficiency are investigated across three basic strategies for automated science: generalization, specialization, and federated specialization. We complement this analysis with numerical screening of information efficiency across strategies.

\section{Measure-Update-Erase Cycles}

Let \(\Theta\) denote the latent environment state with prior \(p_0(\theta)\), Shannon entropy \(H(\Theta_0)=H(p_0)\). At round \(t=1,\dots,T\), the agent chooses an intervention/control \(u_t\) (adapted to the history), obtains an outcome \(Y_t\sim p(y_t\mid \theta, u_t)\), and updates its belief to \(p_t(\theta)=p(\theta\mid y_{1:t},u_{1:t})\). One can write the mutual information contributed at round \(t\) (conditioned on the history) as

\begin{equation}
I_t \;\equiv\; I(\Theta; Y_t \mid Y_{1:t-1}, U_{1:t}) .
\end{equation}

The cumulative information acquired after \(T\) rounds is
\begin{equation}
I_{1:\tau} \;=\; \sum_{t=1}^{\tau} I_t \;=\; H(\Theta_0) - \langle H(\Theta_\tau)\rangle \;\le\; H(\Theta_0).
\label{eq:telescoping}
\end{equation}

In other words, the total information gained is equal to the expected total removal of uncertainty from \(\Theta\).

Each round is defined as a \emph{measure–update–erase} cycle on a working memory or register. Following Sagawa-Ueda and Landauer \cite{Sagawa2010,Landauer1961}, the minimal average work required to perform a measurement and to subsequently erase the measurement record is bounded by
\begin{align}
\langle W_{\mathrm{meas},t}\rangle &\;\ge\; \Delta F_{\mathrm{mem},t} + k_B T\, I_t,
\label{eq:meas-step}\\
\langle W_{\mathrm{erase},t}\rangle &\;\ge\; k_B T\, H(Y_t^{\mathrm{mem}}).
\label{eq:erase-step}
\end{align}
Here, $\Delta F_{\mathrm{mem},t}$ is the change in Helmholtz free energy of the physical memory used to store the outcome at round $t$, and $Y_t^{\mathrm{mem}}$ denotes the logical state of that memory (which may be a discretized representation of the raw outcome $Y_t$). Equation~\eqref{eq:erase-step} is the Landauer bound for erasing a memory with Shannon entropy $H(Y_t^{\mathrm{mem}})$.

In the canonical Sagawa-Ueda setting, the measurement step increases the memory free energy by $\Delta F_{\mathrm{mem},t}\ge 0$, and erasure returns the memory to its reference equilibrium. In this work, we consider an architecture in which the working memory is a small, repeatedly reused register, while the persistent record of each round is stored in a separate, large ``archive'' (for example, a database or model parameters). We idealize this archive as an information reservoir with negligible change in free energy over the $\tau$ rounds, so that the free-energy of the small working register obeys
\begin{equation}
\Delta F_{\mathrm{mem},t} \approx 0
\quad\Rightarrow\quad
\langle W_{\mathrm{meas},t}\rangle \;\gtrsim\; k_B T\, I_t.
\label{eq:meas-simplified}
\end{equation}
More generally, one can interpret \eqref{eq:meas-simplified} as a bound on the excess work above the reversible free-energy change of the memory. The inequalities below therefore hold under a standard idealization in that they underestimate the total energetic cost whenever $\Delta F_{\mathrm{mem},t}>0$, and they should be read as best-case baselines rather than tight bounds for a particular hardware implementation.

Therefore, the total work in round \(t\) obeys
\begin{equation}
\langle W_t\rangle \;=\; \langle W_{\mathrm{meas},t}\rangle + \langle W_{\mathrm{erase},t}\rangle 
\;\ge\; k_B T \,\big(I_t + H(Y_t)\big).
\label{eq:round-bound}
\end{equation}

The work cost in \eqref{eq:round-bound} is written in terms of $H(Y_t)$ for simplicity, implicitly assuming that the agent stores the full outcome as its logical memory state. In general, the agent could first compress $Y_t$ to a statistic $S_t = g(Y_t)$ before committing it to memory. In that case the relevant entropy in \eqref{eq:erase-step} would be $H(S_t)$, and the redundancy $H(Y_t) - H(S_t)$ would not contribute to the erasure cost. From an information-theoretic perspective, the information gain is $I_t = I(\Theta;Y_t\mid Y_{1:t-1},U_{1:t})$, whereas the thermodynamic cost depends on the entropy of whatever representation is actually stored. Our choice to work with $H(Y_t)$ therefore yields a conservative, worst-case baseline. Any compression of $Y_t$ into a lower-entropy representation can only increase the achievable efficiency $\eta_\tau$ relative to the bounds we show below.

Summing over \(\tau\) rounds yields
\begin{equation}
W_{\mathrm{tot}}\;\equiv\;\sum_{t=1}^{\tau} \langle W_t\rangle \;\ge\; k_B T \sum_{t=1}^{\tau} \big(I_t + H(Y_t)\big).
\label{eq:sum-rounds}
\end{equation}

\paragraph{Information-work efficiency.}
For any strategy, define the scale-free information-work efficiency after \(\tau\) rounds and total invested work \(W_{\mathrm{tot}}\):
\begin{equation}
\eta_T \;\equiv\; \frac{I_{1:\tau}}{\beta\, W_{\mathrm{tot}}}\,, \qquad \beta \equiv (k_B T)^{-1}.
\label{eq:eta-def}
\end{equation}

Combining \eqref{eq:sum-rounds} with \eqref{eq:eta-def}, and recognizing the nonnegativity of outcome entropy \(H(Y_t)\), gives the efficiency upper bound
\begin{equation}
\eta_{\tau} \;\le\; \frac{\sum_{t=1}^{\tau} I_t}{\sum_{t=1}^{\tau}\big(I_t + H(Y_t)\big)} 
\;\le\; 1,
\label{eq:eta-upper-entropy}
\end{equation}
with equality only in the unphysical limit of both vanishing outcome entropy and a reversible measurement.

\section{Learning Under a Finite Work Budget}
\paragraph{Finite budget, finite rounds.} Assume a fixed total work budget \(W_{\mathrm{tot}}\) allocated across \(\tau\) rounds (stopping when the budget is exhausted), with \(W_t\) expended each round:
\begin{equation}
    W_\mathrm{tot} - \sum_{t=1}^{\tau}W_t=0
\end{equation}

We can derive \(\tau\) by considering the average work expended by the agent each round across all \(\tau\) rounds:
\begin{equation}
    \frac{W_\mathrm{tot}}{\tau}=\frac{1}{\tau}\sum_{t=1}^{\tau}W_t=\langle W_t\rangle \Rightarrow \tau = \frac{W_\mathrm{tot}}{\langle W_t\rangle}
\end{equation}

\paragraph{Efficiency bound.} Starting from \eqref{eq:sum-rounds}:
\begin{equation}
\beta\, W_{\mathrm{tot}} \;\ge\; \sum_{t=1}^{\tau} \big(I_t + H(Y_t)\big) \;\;\Rightarrow\;\; I_{1:\tau} \;\le\; \beta W_{\mathrm{tot}} - \sum_{t=1}^{T} H(Y_t).
\label{eq:budget-upper}
\end{equation}

Together with \eqref{eq:telescoping} we obtain the fundamental budget–information tradeoff for any unpartitioned strategy:
\begin{equation}
I_{1:\tau} \;\le\; \min\!\Big\{\, H(\Theta_0)\;,\;\beta W_{\mathrm{tot}} - \sum_{t=1}^{\tau} H(Y_t)\,\Big\}.
\label{eq:unpart-I-upper}
\end{equation}

The corresponding efficiency bound follows by dividing \eqref{eq:unpart-I-upper} by \(\beta W_{\mathrm{tot}}\):
\begin{equation}
\eta_\tau^{\mathrm{(unpart)}} \;=\; \frac{I_{1:\tau}}{\beta W_{\mathrm{tot}}}
\;\le\; \min\!\left\{ \frac{H(\Theta_0)}{\beta W_{\mathrm{tot}}},\; 1 - \frac{\sum_{t=1}^{\tau} H(Y_t)}{\beta W_{\mathrm{tot}}} \right\}.
\label{eq:unpart-eta-upper}
\end{equation}

Equation \eqref{eq:unpart-eta-upper} separates two regimes:
\begin{enumerate}
\item \textbf{Prior-limited:} \(\beta W_{\mathrm{tot}} \gg H(\Theta_0)\) \(\Rightarrow\) \(I_{1:\tau}\) saturates at \(H(\Theta_0)\) (no more learnable information remains).
\item \textbf{Budget-limited:} \(\beta W_{\mathrm{tot}} \ll H(\Theta_0)\) \(\Rightarrow\) \(I_{1:\tau}\) is capped by the work budget and further reduced by the sum of the outcome entropies.
\end{enumerate}

\section{Partitioning the Environment}
\label{sec:federated}

A federated strategy partitions the environment into $N$ subdomains, indexed by a discrete random variable $K\in\{1,\dots,N\}$ with prior masses
\[
p_i \;\equiv\; \Pr[K=i], \qquad \sum_{i=1}^N p_i = 1.
\]
Conditioned on $K=i$, the environment is described by the conditional prior $p(\theta \mid K=i)$ over the same underlying state $\Theta_0$ as the unpartitioned strategy. One can denote this conditional prior as $\Theta_0^{(i)} \sim p^{(i)}_0(\theta)$ and write its entropy as
\[
H_i \;\equiv\; H(\Theta_0^{(i)}) \;=\; H(\Theta_0 \mid K=i).
\]
The probabilities $p_i$ encode how likely it is a priori that the informational objective \(\Theta_0\) lies in subdomain $i$. Within each subdomain, the conditional distributions $p^{(i)}_0(\theta)$ represent the remaining uncertainty about the state \(\Theta_0\) given that the relevant information is in that subdomain. Thus, the federated description carries a two-layer Bayesian structure: a prior over subdomains $(p_i)$ and, a prior over states $(p^{(i)}_0)$ which is conditioned on each subdomain.

A partitioned (federated) round consists of a \emph{measure-update–erase} cycle carried out by the subagent responsible for the realized subdomain $K$.
Over the course of $\tau$ rounds, subagent $i$ uses a total work budget $W_{\mathrm{tot}}^{(i)}$, with the global budget constrained by
\[
W_{\mathrm{tot}} \;=\; \sum_{i=1}^N p_i\, W_{\mathrm{tot}}^{(i)},
\]
so that $W_{\mathrm{tot}}$ is the expected total work across subdomains, weighted by how frequently each subdomain is engaged.
At round $t$, when $K=i$ is active, we denote by
\[
I_{i,t} \;\equiv\; I\!\big(\Theta_0;Y_{i,t}\mid Y_{1:t-1}, U_{1:t}, K=i\big),
\qquad
H(Y_{i,t}) \;\equiv\; H\!\big(Y_{i,t}\mid K=i\big),
\]
the mutual information and outcome entropy within subdomain $i$.

\paragraph{Federated efficiency bound.}
For each fixed subdomain $i$, the single-subdomain \emph{measure–update–erase} cycle applied to the $\tau$ rounds in which $K=i$ yields
\begin{equation}
\sum_{t=1}^{\tau} I_{i,t}
\;\le\;
\min\!\Big\{
H(\Theta_0^{(i)}),\;
\beta W_{\mathrm{tot}}^{(i)} - \sum_{t=1}^{\tau} H(Y_{i,t})
\Big\}.
\label{eq:per-subdomain}
\end{equation}
Define the expected total information gain under the federated strategy as
\[
I^{\mathrm{(fed)}}_{1:\tau}
\;\equiv\;
\mathbb{E}_K\!\left[\sum_{t=1}^{\tau} I_{K,t}\right]
\;=\;
\sum_{i=1}^{N} p_i \sum_{t=1}^{\tau} I_{i,t}.
\]
Multiplying \eqref{eq:per-subdomain} by $p_i$ and summing over $i$ gives
\begin{align}
I^{\mathrm{(fed)}}_{1:\tau}
&\le
\sum_{i=1}^{N} p_i
\min\!\Big\{
H(\Theta_0^{(i)}),\;
\beta W_{\mathrm{tot}}^{(i)} - \sum_{t=1}^{\tau} H(Y_{i,t})
\Big\}.
\label{eq:fed-I-upper}
\end{align}
We define the federated information–work efficiency over $\tau$ rounds as
\[
\eta_\tau^{\mathrm{(fed)}} \;\equiv\; \frac{I^{\mathrm{(fed)}}_{1:\tau}}{\beta W_{\mathrm{tot}}}.
\]
Using \eqref{eq:fed-I-upper} and the fact that $\min\{x,y\}\le x$ and $\min\{x,y\}\le y$ for any $x,y$, we obtain two global caps:
\begin{align*}
I^{\mathrm{(fed)}}_{1:\tau}
&\le
\sum_{i=1}^{N} p_i\, H(\Theta_0^{(i)})
\;=\;
H(\Theta_0 \mid K),
\\[4pt]
I^{\mathrm{(fed)}}_{1:\tau}
&\le
\sum_{i=1}^{N} p_i
\Big(
\beta W_{\mathrm{tot}}^{(i)} - \sum_{t=1}^{\tau} H(Y_{i,t})
\Big)
\;=\;
\beta W_{\mathrm{tot}}
-
\sum_{i=1}^{N} p_i \sum_{t=1}^{\tau} H(Y_{i,t}).
\end{align*}
Combining these, the federated strategy obeys
\begin{align}
\eta_\tau^{\mathrm{(fed)}}
&= \frac{I^{\mathrm{(fed)}}_{1:\tau}}{\beta W_{\mathrm{tot}}}
\notag\\[4pt]
&\le
\min\!\left\{
\frac{H(\Theta_0 \mid K)}{\beta W_{\mathrm{tot}}},\;
1 - \frac{1}{\beta W_{\mathrm{tot}}}
\sum_{i=1}^{N} p_i \sum_{t=1}^{\tau} H(Y_{i,t})
\right\}.
\label{eq:fed-eta-upper}
\end{align}
It is convenient to define
\[
H_{\mathrm{gen}} \;\equiv\; H(\Theta_0),
\qquad
H_{\mathrm{fed}} \;\equiv\; H(\Theta_0\mid K)
\;=\;
\sum_{i=1}^{N} p_i\, H(\Theta_0^{(i)}),
\]
so that \eqref{eq:fed-eta-upper} resembles the unpartitioned bound \eqref{eq:unpart-eta-upper} with $H(\Theta_0)$ replaced by $H_{\mathrm{fed}}$.

\paragraph{Entropic benefits of partitioning.}
By the chain rule for entropy,
\begin{equation}
H_{\mathrm{fed}}
\;=\;
H(\Theta_0\mid K)
\;=\;
H(\Theta_0) - I(\Theta_0;K)
\;=\;
H_{\mathrm{gen}} - I(\Theta_0;K)
\;\le\;
H_{\mathrm{gen}},
\label{eq:Hfed-less-Hgen}
\end{equation}
with strict inequality whenever the partition is informative $(I(\Theta_0;K)>0)$.
Thus, any informative partitioning of the environment strictly reduces the entropy term in the efficiency bound, replacing $H_{\mathrm{gen}}$ by a lower $H_{\mathrm{fed}}$ in the prior-limited regime.
In a heterogeneous domain with sparse outer-product structure (suggesting a highly non-uniform density of state), a coarse partition $K$ can remove large combinatorial density, so $I(\Theta_0;K)$ can be large and $H_{\mathrm{fed}}\ll H_{\mathrm{gen}}$ in practice.

\section{Strategies at the Limits of K-Partitions}

The $K$-based formulation contains both the unpartitioned generalist and the single-domain specialist strategies as limiting cases.

\paragraph{Generalist as a trivial partition.}
If there is no meaningful partition, we can regard $K$ as a trivial random variable with a single value, say $K\equiv 1$. The prior over $K$ is then
\[
p_1 = 1, \qquad p_i = 0 \;\; (i\neq 1),
\]
and the conditional prior over $\Theta_0$ given $K=1$ coincides with the original prior:
\[
\Theta_0^{(1)} \sim p(\theta \mid K=1) = p_0(\theta), 
\qquad
H(\Theta_0^{(1)}) = H(\Theta_0) \equiv H_{\mathrm{gen}}.
\]
All work is spent in this single branch, so
\[
W_{\mathrm{tot}}^{(1)} = W_{\mathrm{tot}}, 
\qquad
W_{\mathrm{tot}}^{(i)} = 0 \;\; (i\neq 1),
\]
and the outcomes and information gains coincide with the unpartitioned notation,
\[
Y_{1,t} = Y_t, 
\qquad 
I_{1,t} = I_t.
\]
Applying the per-subdomain bound \eqref{eq:per-subdomain} to $i=1$ gives
\[
\sum_{t=1}^{\tau} I_{1,t}
\;\le\;
\min\!\Big\{
H(\Theta_0^{(1)}),\;
\beta W_{\mathrm{tot}}^{(1)} - \sum_{t=1}^{\tau} H(Y_{1,t})
\Big\}
=
\min\!\Big\{
H_{\mathrm{gen}},\;
\beta W_{\mathrm{tot}} - \sum_{t=1}^{\tau} H(Y_t)
\Big\}.
\]
Since $p_1=1$, the federated total information gain is
\[
I^{\mathrm{(fed)}}_{1:\tau}
=
\sum_{i=1}^N p_i \sum_{t=1}^{\tau} I_{i,t}
=
\sum_{t=1}^{\tau} I_{1,t}
\equiv
I_{1:\tau},
\]
and the federated efficiency reduces to
\[
\eta_\tau^{\mathrm{(fed)}}
=
\frac{I^{\mathrm{(fed)}}_{1:\tau}}{\beta W_{\mathrm{tot}}}
=
\frac{I_{1:\tau}}{\beta W_{\mathrm{tot}}}
\;\le\;
\min\!\left\{
\frac{H_{\mathrm{gen}}}{\beta W_{\mathrm{tot}}},\;
1 - \frac{\sum_{t=1}^{\tau} H(Y_t)}{\beta W_{\mathrm{tot}}}
\right\},
\]
which is exactly the unpartitioned bound \eqref{eq:unpart-eta-upper}. In this sense, the \textbf{generalist} strategy is a trivial partition with $N=1$, $H_{\mathrm{fed}} = H(\Theta_0\mid K) = H_{\mathrm{gen}}$, and no reduction in prior entropy.

\paragraph{Specialist as a concentrated partition.}
A single \textbf{specialist} that focuses all work on one subdomain $i^\star$ is recovered by taking a nontrivial partition $K \in \{1,\dots,N\}$ but a degenerate prior
\[
p_{i^\star} = 1,\qquad p_i = 0 \;\; (i\neq i^\star),
\]
and allocating the full budget to that branch, $W_{\mathrm{tot}}^{(i^\star)} = W_{\mathrm{tot}}$,
$W_{\mathrm{tot}}^{(i)}=0$ for $i\neq i^\star$.
In this case,
\[
H_{\mathrm{fed}}
= H(\Theta_0 \mid K)
= H(\Theta_0 \mid K=i^\star)
\equiv H_{\mathrm{spec}},
\]
and the federated bound \eqref{eq:fed-eta-upper} becomes
\[
\eta_\tau^{\mathrm{(spec)}}
\;\le\;
\min\!\left\{
\frac{H_{\mathrm{spec}}}{\beta W_{\mathrm{tot}}},\;
1 - \frac{1}{\beta W_{\mathrm{tot}}}
\sum_{t=1}^{\tau} H(Y_{i^\star,t})
\right\},
\]
which is just the single-domain \emph{measure–update–erase} bound applied to the specialist's subdomain.

In summary, the partition variable $K$ provides a unified description in which
(i) the generalist corresponds to a trivial partition with $H_{\mathrm{fed}} = H_{\mathrm{gen}}$, (ii) a single specialist is a degenerate partition that concentrates $p(K)$ and work on one subdomain, and (iii) a fully federated strategy uses a non-degenerate $p(K)$ to spread work across multiple subdomains, reducing the relevant entropy from $H_{\mathrm{gen}}$ to $H_{\mathrm{fed}} = H(\Theta_0 \mid K)$ in the prior-limited regime.

\section{Numerical Screening of Information–Work Efficiency}

The analytic bounds developed above show that architectural differences between strategies enter through two thermodynamic channels: (i) the effective prior entropy (e.g.\ $H_{\mathrm{gen}}$, $H_{\mathrm{fed}}$), and (ii) the outcome entropy $H(Y)$ appearing in the measurement–erasure work cost. Here we use a simple toy model to screen how these two levers jointly shape the information–work efficiency of generalist, federated, and specialist architectures across regimes of total work budget and partition granularity.

\subsection{Toy efficiency model}
We normalize work by the generalist prior entropy and work in terms of the dimensionless budget
\begin{equation}
\omega \;\equiv\; \frac{\beta W_{\mathrm{tot}}}{H_{\mathrm{gen}}}.
\end{equation}
For each strategy $\mathrm{str}\in\{\mathrm{gen},\mathrm{fed},\mathrm{spec}\}$ we posit a minimal toy efficiency law
\begin{equation}
\eta^{(\mathrm{str})}(\omega)
= \min\!\left\{
\frac{c_{\mathrm{str}}}{\omega},
\;
\frac{1}{1+\alpha_{\mathrm{str}}}
\right\},
\label{eq:numeric-eta}
\end{equation}
where:
\begin{itemize}
\item $c_{\mathrm{str}} \equiv H_{\mathrm{eff}}^{(\mathrm{str})}/H_{\mathrm{gen}}$ is an effective prior-entropy ratio for that strategy
\item $\alpha_{\mathrm{str}}$ is a dimensionless measure of the outcome-entropy contribution $\sum_t H(Y_t)$ and irreversibility penalty, normalized so that $1/(1+\alpha_{\mathrm{str}})$ plays the role of an overhead ceiling in the budget-limited regime.
\end{itemize}
The generalist faces the full prior entropy, so $c_{\mathrm{gen}}=1$. For a specialist focused on a single subdomain we set
\[
c_{\mathrm{spec}} \in [c_{\min},1], \qquad
c_{\mathrm{spec}} = \frac{H_{\mathrm{spec}}}{H_{\mathrm{gen}}},
\]
where $c_{\min}\ll 1$ corresponds to an extremely narrow niche. For a federated strategy with $N$ subdomains we model the compression of effective entropy as
\begin{equation}
c_{\mathrm{fed}}(N)
\;=\;
\frac{H_{\mathrm{fed}}(N)}{H_{\mathrm{gen}}}
\;=\;
c_{\min} + \frac{1-c_{\min}}{N^\gamma},
\label{eq:c-fed}
\end{equation}
where $\gamma>0$ controls how quickly increasing $N$ reduces the effective prior entropy. In the limit $N\to 1$ we recover $c_{\mathrm{fed}}\to 1$ (a generalist), and for large $N$ we approach $c_{\mathrm{fed}}\to c_{\min}$ (a very fine federated decomposition).

Within this model, a strategy is more efficient when it either (i) reduces effective prior entropy (lower $c_{\mathrm{str}}$), or (ii) reduces measurement/erasure overhead (lower $\alpha_{\mathrm{str}}$). We visualize these tradeoffs via pairwise efficiency differences
\begin{align}
\Delta\eta_{\mathrm{spec-gen}} &= \eta_{\mathrm{spec}} - \eta_{\mathrm{gen}},\\
\Delta\eta_{\mathrm{fed-gen}} &= \eta_{\mathrm{fed}} - \eta_{\mathrm{gen}},\\
\Delta\eta_{\mathrm{fed-spec}} &= \eta_{\mathrm{fed}} - \eta_{\mathrm{spec}},
\end{align}
across grids in $(\omega,c_{\mathrm{spec}})$ and $(\omega,N)$. Regions with $\Delta\eta>0$ favor the first strategy, and regions with $\Delta\eta<0$ favor the second. The black curves in Fig.~\ref{fig:delta-eta-phase} mark $\Delta\eta=0$ contours: phase boundaries where the two strategies are equally efficient.

For the numerical experiments in Fig.~\ref{fig:delta-eta-phase}, we fix units such that $\beta = 1$ and $H_{\mathrm{gen}} = 1$, so the normalized budget is simply $\omega = W_{\mathrm{tot}}$. We take
$c_{\min} = 0.05$ and $\gamma = 1.0$ in the federated compression law $c_{\mathrm{fed}}(N) = c_{\min} + (1-c_{\min})/N^\gamma$, so that $c_{\mathrm{fed}}(1) = 1$ (a generalist) and $c_{\mathrm{fed}}(N)\to c_{\min}$ for large $N$. The grids used in the heatmaps are $\omega \in [10^{-2},10^{2}]$ (log-spaced), $c_{\mathrm{spec}} \in [c_{\min},1]$, and $N \in [1,20]$ (treated as continuous for visualization). In the symmetric outcome-entropy regime we set a common value $\alpha_{\mathrm{gen}} = \alpha_{\mathrm{fed}} = \alpha_{\mathrm{spec}} \equiv \alpha_{\mathrm{sym}} = 0.3$. In the asymmetric regime, we choose $(\alpha_{\mathrm{gen}},\alpha_{\mathrm{fed}},\alpha_{\mathrm{spec}}) = (0.8,0.4,0.2)$ to model a high-overhead generalist, an intermediate federated architecture, and a more reversible specialist.

\subsection{Symmetric outcome-entropy regime}
We first consider a symmetric regime in which all three strategies incur the same outcome-entropy penalty
\[
\alpha_{\mathrm{gen}} = \alpha_{\mathrm{fed}} = \alpha_{\mathrm{spec}} \equiv \alpha_{\mathrm{sym}},
\]
so strategy can only change efficiency through the prior-entropy channel $c_{\mathrm{str}}$.

Panels~A, C, and E in Fig.~\ref{fig:delta-eta-phase} show $\Delta\eta$ surfaces in this symmetric setting. The vertical white dashed line at $\omega\simeq 1$ marks the approximate transition between the budget-limited regime ($\omega\ll 1$) and the prior-limited regime ($\omega\gg 1$):

\begin{itemize}
\item In the \textbf{budget-limited regime} ($\omega\ll 1$), all strategies are capped by the same overhead ceiling $1/(1+\alpha_{\mathrm{sym}})$, so differences in $c_{\mathrm{str}}$ do not yet matter. The heatmaps are nearly flat in this region and the black $\Delta\eta=0$ curves (where present) parallel the vertical axis.

\item In the \textbf{prior-limited regime} ($\omega\gg 1$), Eq.~\eqref{eq:numeric-eta} reduces to $\eta^{(\mathrm{str})} \approx c_{\mathrm{str}}/\omega$. Here, lower effective prior entropy helps, but because $c_{\mathrm{gen}}=1$ is the largest possible value, the generalist maintains the highest efficiency: the symmetric simulations produce a consistent ordering
\[
\eta_{\mathrm{gen}}(\omega) \;\ge\; \eta_{\mathrm{fed}}(\omega) \;\ge\; \eta_{\mathrm{spec}}(\omega)
\quad
\text{for all } \omega,
\]
with the federated strategy sitting between the generalist and the maximally focused specialist. Panels~C and E show that while federation can outperform a single extremely focused specialist (green regions with $\Delta\eta_{\mathrm{fed-spec}}>0$), neither strategy ever beats the generalist when outcome entropy is held fixed.
\end{itemize}

Thus, purely reducing effective prior entropy while keeping $H(Y)$ unchanged is insufficient to make federated specialization more efficient than a generalist under matched budgets.

\subsection{Asymmetric outcome-entropy regime}

Real architectures are unlikely to be symmetric in outcome entropy. Global, multiplexed generalist systems often require high-bandwidth control, noisy interventions, and complex erasure steps, all of which inflate $H(Y)$ and the associated thermodynamic overhead. Domain-specific modules, by contrast, can be engineered to be more reversible and structured. To capture this, we study an asymmetric regime in which
\[
\alpha_{\mathrm{gen}} > \alpha_{\mathrm{fed}} > \alpha_{\mathrm{spec}},
\]
so specialists pay the smallest overhead, federated modules are intermediate, and the generalist pays the highest outcome-entropy cost.

\begin{figure}[H]
    \centering
    \includegraphics[width=\linewidth]{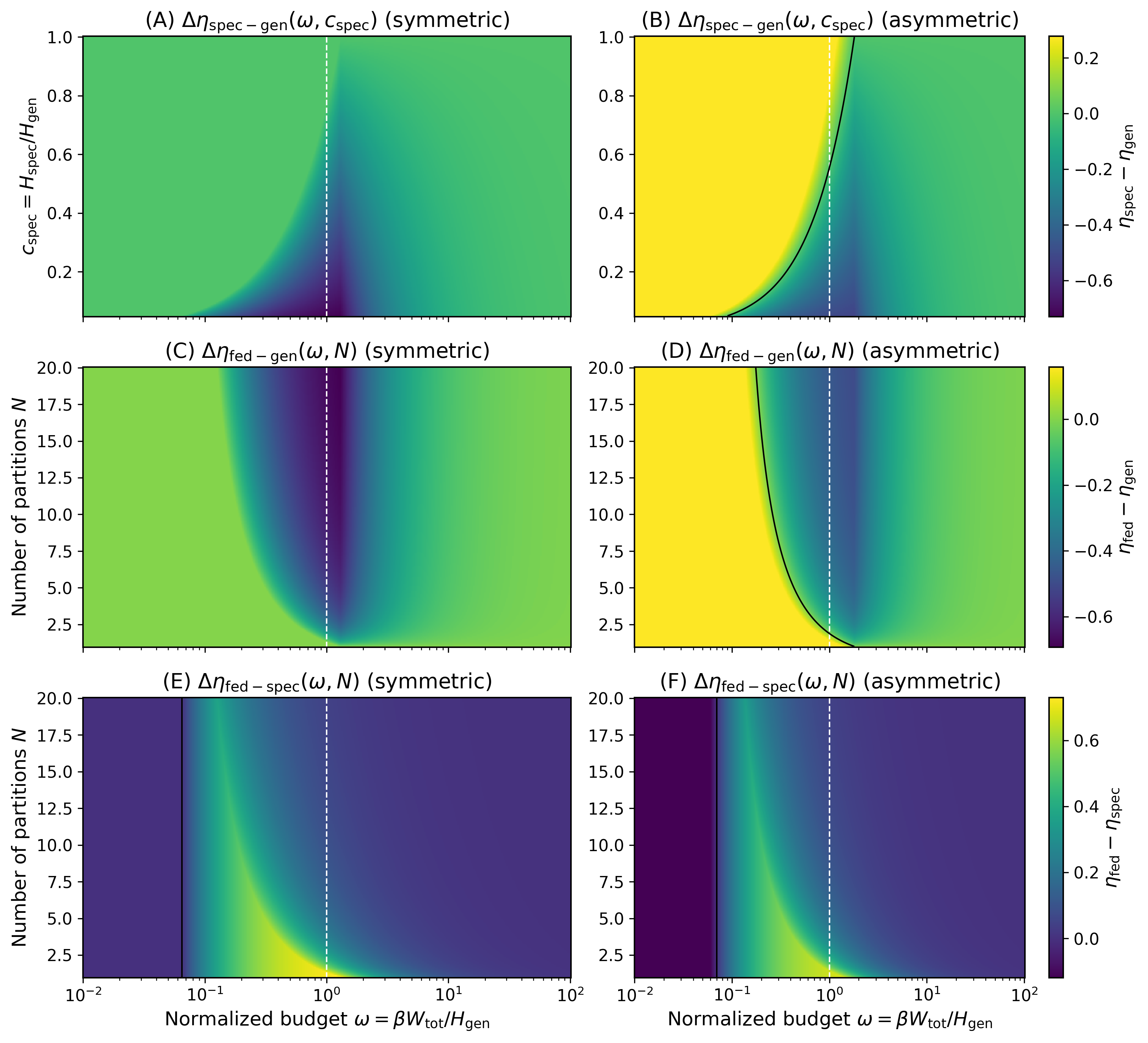}
    \caption{\textbf{Pairwise efficiency differences across strategies.}
    Each panel shows a heatmap of a pairwise efficiency difference $\Delta\eta$ under the toy model~\eqref{eq:numeric-eta} as a function of normalized work budget $\omega = \beta W_{\mathrm{tot}}/H_{\mathrm{gen}}$ and either specialization level $c_{\mathrm{spec}}$ (top row) or number of partitions $N$ (middle and bottom rows).
    Left column: symmetric outcome-entropy regime with $\alpha_{\mathrm{gen}}=\alpha_{\mathrm{fed}}=\alpha_{\mathrm{spec}}$.
    Right column: asymmetric regime with $\alpha_{\mathrm{gen}}>\alpha_{\mathrm{fed}}>\alpha_{\mathrm{spec}}$.
    (A,B) Specialist vs.\ generalist, $\Delta\eta_{\mathrm{spec-gen}}$.
    (C,D) Federated vs.\ generalist, $\Delta\eta_{\mathrm{fed-gen}}$.
    (E,F) Federated vs.\ maximally focused specialist, $\Delta\eta_{\mathrm{fed-spec}}$ (with $c_{\mathrm{spec}}=c_{\min}$).
    The vertical white dashed line marks $\omega\simeq 1$, separating budget-limited (left) and prior-limited (right) regions.
    Black curves indicate the $\Delta\eta=0$ contours, i.e.\ phase boundaries where the two strategies are equally efficient.
    Yellow regions favor the first strategy (positive $\Delta\eta$), purple regions favor the second.}
    \label{fig:delta-eta-phase}
\end{figure}

Panels~B, D, and F in Fig.~\ref{fig:delta-eta-phase} show the corresponding $\Delta\eta$ landscapes. Several qualitative features emerge:
\begin{itemize}
\item In the \textbf{budget-limited} region ($\omega\ll 1$), all strategies are pinned near their respective ceilings $1/(1+\alpha_{\mathrm{str}})$, and the ordering is governed by overhead alone. This produces wide yellow bands in panels~B and~D where $\Delta\eta_{\mathrm{spec-gen}}>0$ and $\Delta\eta_{\mathrm{fed-gen}}>0$: at sufficiently low budgets, both specialists and federated systems are more efficient than the generalist simply because they are more thermodynamically reversible.

\item As we move into the \textbf{prior-limited} region ($\omega\gg 1$), the entropy-compression factor $c_{\mathrm{str}}$ becomes increasingly important. In panel~D, the black $\Delta\eta_{\mathrm{fed-gen}}=0$ curve bends to the right with increasing $N$, separating a low-$\omega$/moderate-$N$ region where the federated strategy dominates (yellow) from a high-$\omega$/small-$N$ region where the generalist regains the advantage (purple). This is exactly the tradeoff predicted by the analytic bounds: federated specialization wins when it can combine (i) reduced prior entropy $c_{\mathrm{fed}}(N)<1$ with (ii) a lower outcome-entropy penalty $\alpha_{\mathrm{fed}}<\alpha_{\mathrm{gen}}$.

\item Panel~F compares federation to a maximally focused specialist ($c_{\mathrm{spec}}=c_{\min}$). At low budgets, the specialist’s lower overhead makes it more efficient (purple region). As $\omega$ and $N$ increase, federation catches up and eventually overtakes the specialist (yellow region), since it retains lower overhead than the generalist while spreading work across multiple subdomains.
\end{itemize}

Taken together, the numerical screening makes the main lesson from the analytic bounds concrete. If partitioning only reshapes prior entropy, the generalist remains optimal. Federated specialization becomes thermodynamically advantageous when it simultaneously reduces the effective entropy of the domain and the outcome entropy of the measurement process. Under those conditions, the black $\Delta\eta=0$ curves in Fig.~\ref{fig:delta-eta-phase} trace out broad regions of $(\omega,N,c_{\mathrm{spec}})$ where federated architectures outperform both a generalist and any single specialist at matched work budgets.

\section*{Conclusion}
We have used a simple \emph{measure–update–erase} model to relate information acquisition in automated science to established bounds in information thermodynamics. This leads to a finite-budget information–work efficiency $\eta$ and explicit upper bounds that separate a prior-limited regime, where learning is constrained by the entropy of the environment, from a budget-limited regime, where learning is constrained by the total work and the outcome entropy of the measurement process. Introducing a partition variable $K$ shows how strategic choices enter the bounds through two quantities: the effective prior entropy $H(\Theta_0\mid K)$ and the aggregated outcome entropy $\sum_t H(Y_t)$. In this description, generalist, specialist, and federated strategies are recovered as limiting cases of the same framework. The analytic bounds show that informative partitioning can reduce the entropy term in the prior-limited regime. Numerical screening highlights these effects in a minimal toy model. When different strategies have the same outcome-entropy penalty, the generalist is never outperformed at matched work budgets, even if federation or specialization reduce effective prior entropy. When the generalist incurs higher outcome entropy, federated strategies can achieve higher efficiency over a range of budgets and partition sizes. Cumulatively, the results provide a way to reason about how strategy behind automated science platforms trade off prior compression and thermodynamic overhead, and indicate which combinations of these factors are necessary for federated designs to gain an efficiency advantage.

Several idealizations underlie our analysis. We have treated the memory architecture in a best-case manner, effectively setting $\Delta F_{\mathrm{mem},t}\approx 0$ for a small working register and taking the stored record to be the full outcome $Y_t$; in practice, both nonzero free-energy changes and lossless compression of $Y_t$ into lower-entropy statistics will shift the achievable efficiencies relative to our baseline bounds. Likewise, we have modeled federated and specialized architectures at the level of epistemic entropies and outcome entropies, rather than committing to a particular physical implementation. These simplifications allow us to isolate the information-theoretic structure of the tradeoffs. Extending the framework to specific hardware and experimental pipelines is a natural direction for future work.


\begin{thebibliography}{9}

\bibitem{Landauer1961}
R. Landauer, ``Irreversibility and Heat Generation in the Computing Process,'' \textit{IBM Journal of Research and Development}, 5(3), 183-191 (1961).
DOI: \href{https://doi.org/10.1147/rd.53.0183}{10.1147/rd.53.0183}.

\bibitem{Bennett1982}
C. H. Bennett, ``The Thermodynamics of Computation-A Review,'' \textit{International Journal of Theoretical Physics}, 21(12), 905-940 (1982).
DOI: \href{https://doi.org/10.1007/BF02084158}{10.1007/BF02084158}.

\bibitem{Sagawa2010}
T. Sagawa and M. Ueda, ``Generalized Jarzynski Equality under Nonequilibrium Feedback Control,'' \textit{Physical Review Letters}, 104(9), 090602 (2010).
DOI: \href{https://doi.org/10.1103/PhysRevLett.104.090602}{10.1103/PhysRevLett.104.090602}.

\bibitem{Jarzynski2011}
C. Jarzynski, ``Equalities and Inequalities: Irreversibility and the Second Law of Thermodynamics at the Nanoscale,'' \textit{Annual Review of Condensed Matter Physics}, 2(1), 329-351 (2011).
DOI: \href{https://doi.org/10.1146/annurev-conmatphys-062910-140506}{10.1146/annurev-conmatphys-062910-140506}.

\bibitem{Toyabe2010}
S. Toyabe, T. Sagawa, M. Ueda, E. Muneyuki, and M. Sano,
``Experimental demonstration of information-to-energy conversion and validation of the generalized Jarzynski equality,''
\textit{Nature Physics}, 6, 988-992 (2010).
DOI: \href{https://doi.org/10.1038/nphys1821}{10.1038/nphys1821}.

\bibitem{Still2012}
S. Still, D. A. Sivak, A. J. Bell, and G. E. Crooks,
``Thermodynamics of prediction,''
\textit{Physical Review Letters}, 109(12), 120604 (2012).
DOI: \href{https://doi.org/10.1103/PhysRevLett.109.120604}{10.1103/PhysRevLett.109.120604}.

\bibitem{Horowitz2014}
J. M. Horowitz and M. Esposito,
``Thermodynamics with continuous information flow,''
\textit{Physical Review X}, 4(3), 031015 (2014).
DOI: \href{https://doi.org/10.1103/PhysRevX.4.031015}{10.1103/PhysRevX.4.031015}.

\end{thebibliography}
\end{document}